\documentclass[11pt]{article}
\usepackage{moriond,epsfig,eucal}

\def\be{\begin{equation}}
\def\ee{\end{equation}}
\def\bea{\begin{eqnarray}}
\def\eea{\end{eqnarray}}

\begin{document}
\title{RESULTS FROM AMANDA}

\author{G.C. HILL, FOR THE AMANDA COLLABORATION }

\address{ Physics Department, University of Wisconsin, Madison,
WI 53706, USA \\
 {\rm email: ghill@alizarin.physics.wisc.edu}\\
\vspace{0.5cm}
{\rm AMANDA COLLABORATION:
J.~AHRENS$^{9}$, 
X.~BAI$^{1}$, 
G.~BAROUCH$^{12}$, 
S.W.~BARWICK$^{8}$, 
R.C.~BAY$^{7}$, 
T.~BECKA$^{9}$, 
K.-H.~BECKER$^{2}$, 
D.~BERTRAND$^{3}$, 
A.~BIRON$^{4}$, 
J.~BOOTH$^{8}$, 
O.~BOTNER$^{13}$, 
A.~BOUCHTA$^{4}$, 
M.M.~BOYCE$^{12}$, 
S.~CARIUS$^{5}$, 
A.~CHEN$^{12}$, 
D.~CHIRKIN$^{7}$, 
J.~CONRAD$^{13}$, 
J.~COOLEY$^{12}$, 
C.G.S.~COSTA$^{3}$, 
D.F.~COWEN$^{11}$, 
C. DECLERCQ~$^{3}$,
T.~DEYOUNG$^{12}$, 
P.~DESIATI$^{4}$, 
J.-P.~DEWULF$^{3}$, 
P.~DOKSUS$^{12}$, 
J.~EDSJ\"O$^{14}$, 
P.~EKSTR\"OM$^{14}$, 
T.~FESER$^{9}$, 
T.~GAISSER$^{1}$, 
M.~GAUG$^{4}$, 
L.~GERHARDT$^{8}$,
A.~GOLDSCHMIDT$^{6}$, 
A.~HALLGREN$^{13}$, 
F.~HALZEN$^{12}$, 
K.~HANSON$^{11}$, 
R.~HARDTKE$^{12}$, 
T.~HAUSCHILDT$^{4}$,
M.~HELLWIG$^{9}$, 
P. HERQUET$^{3}$,
G.C.~HILL$^{12}$, 
P.O.~HULTH$^{14}$, 
S.~HUNDERTMARK$^{8}$, 
J.~JACOBSEN$^{6}$, 
A.~KARLE$^{12}$, 
J.~KIM$^{8}$, 
B.~KOCI$^{12}$, 
L.~K\"OPKE$^{9}$, 
M.~KOWALSKI$^{4}$, 
J.I.~LAMOUREUX$^{6}$, 
H.~LEICH$^{4}$, 
M.~LEUTHOLD$^{4}$, 
P.~LINDAHL$^{5}$, 
J.~MADSEN$^{12}$, 
P.~MARCINIEWSKI$^{13}$, 
H.S.~MATIS$^{6}$, 
Y.~MINAEVA$^{14}$, 
P.~MIO\v{C}INOVI\'C$^{7}$, 
R.~MORSE$^{12}$, 
T.~NEUNH\"OFFER$^{9}$, 
P.~NIESSEN$^{4}$, 
D.R.~NYGREN$^{6}$, 
H.~OGELMAN$^{12}$, 
P. OLBRECHTS~$^{3}$, 
C.~P\'EREZ~DE~LOS~HEROS$^{13}$, 
P.B.~PRICE$^{7}$, 
K.~RAWLINS$^{12}$, 
C.~REED$^{8}$, 
M.~RIBORDY$^{4}$, 
W.~RHODE$^{2}$, 
S.~RICHTER$^{4}$, 
J.~RODR\'\i GUEZ~MARTINO$^{14}$, 
D.~ROSS$^{8}$, 
H.-G.~SANDER$^{9}$, 
T.~SCHMIDT$^{4}$, 
D.~SCHNEIDER$^{12}$, 
A.~SILVESTRI$^{2}$, 
M.~SOLARZ$^{7}$, 
G.M.~SPICZAK$^{1}$, 
C.~SPIERING$^{4}$, 
N.~STARINSKY$^{12}$, 
D.~STEELE$^{12}$, 
P.~STEFFEN$^{4}$, 
R.G.~STOKSTAD$^{6}$, 
P. SUDHOFF$^{4}$, 
K.-H. SULANKE$^{4}$, 
I.~TABOADA$^{11}$, 
M.~VANDER~DONCKT$^{3}$, 
C.~WALCK$^{14}$, 
C.~WEINHEIMER$^{9}$, 
C.H.~WIEBUSCH$^{4}$, 
R.~WISCHNEWSKI$^{4}$, 
H.~WISSING$^{4}$, 
K.~WOSCHNAGG$^{7}$, 
G.~YODH$^{8}$, 
S.~YOUNG$^{8}$}\\
\vspace{0.5cm}
1 Bartol Research Institute, University of Delaware, Newark, DE 19716, USA  \\
2 Fachbereich 8 Physik, BUGH Wuppertal, D-42097 Wuppertal, Germany  \\
3 Brussels Free University, Science Faculty CP230, Boulevard du 
Triomphe, B-1050 Brussels, Belgium   \\
4 DESY-Zeuthen, D-15735 Zeuthen, Germany  \\
5 Dept. of Technology, Kalmar University, S-39182 Kalmar, Sweden  \\
6 Lawrence Berkeley National Laboratory, Berkeley, CA 94720, USA  \\
7 Dept. of Physics, University of California, Berkeley, CA 94720, USA  \\
8 Dept. of Physics and Astronomy, University of California, Irvine, CA 92697, USA  \\
9 Institute of Physics, University of Mainz, Staudinger Weg 7, D-55099 Mainz, Germany  \\
10 Dept. of Physics, University of Maryland,   \\
11 Dept. of Physics and Astronomy, University of Pennsylvania, 
Philadelphia, PA 19104, USA  \\
12 Dept. of Physics, University of Wisconsin, Madison, WI 53706, USA  \\
13 Division of High Energy Physics, Uppsala University, S-75121 Uppsala, Sweden  \\
14 Fysikum, Stockholm University, S-11385 Stockholm, Sweden
}

\maketitle

\section{Introduction}

\begin{figure}[htb]
\centering
\mbox{
\epsfig{figure=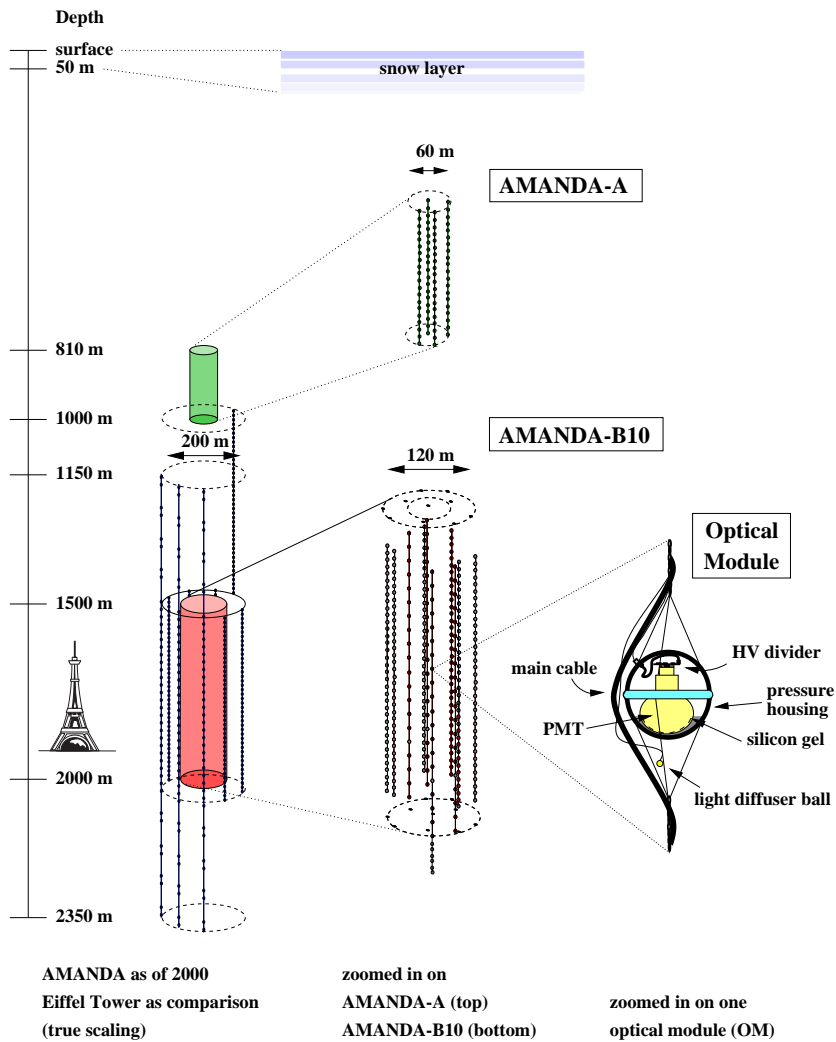,width=11.0cm}
}
\caption{
The AMANDA-II array at the South Pole}
\label{fig:amanda_2}
\end{figure}

The AMANDA (Antarctic Muon And Neutrino Detector Array) detector,
 located at the South Pole
station, Antarctica, was recently expanded 
with the addition of six new strings, completing the phase referred to as
AMANDA-II. This detector has been calibrated and in operation since January 2000.
 Figure \ref{fig:amanda_2} shows a schematic view of the 
new AMANDA-II array.  The first
data analyses are currently underway. In this report we present an update on the 
results from the AMANDA-B10 detector, which operated during the austral winter 1997.
 This detector,
 consisted
of 302 optical sensors on 10 strings located at depths of 1500 to 2000\,m
in the deep Antarctic ice.   

The main science goal of the AMANDA project is to search for extra-terrestrial
neutrinos, from sources such as active galaxies, or Gamma-ray bursts. As a precursor
to such a search, we have focussed on the detection of atmospheric neutrinos, which
essentially act as a calibration source, allowing us to check the correct performance
and understanding of the detector. In this paper, we first discuss the atmospheric
neutrino search in some detail, then report on the status of the search for
 extra-terrestrial neutrinos.

\section{Atmospheric Neutrinos}

\begin{figure}[ht]
\centering
  \mbox{\epsfig{file=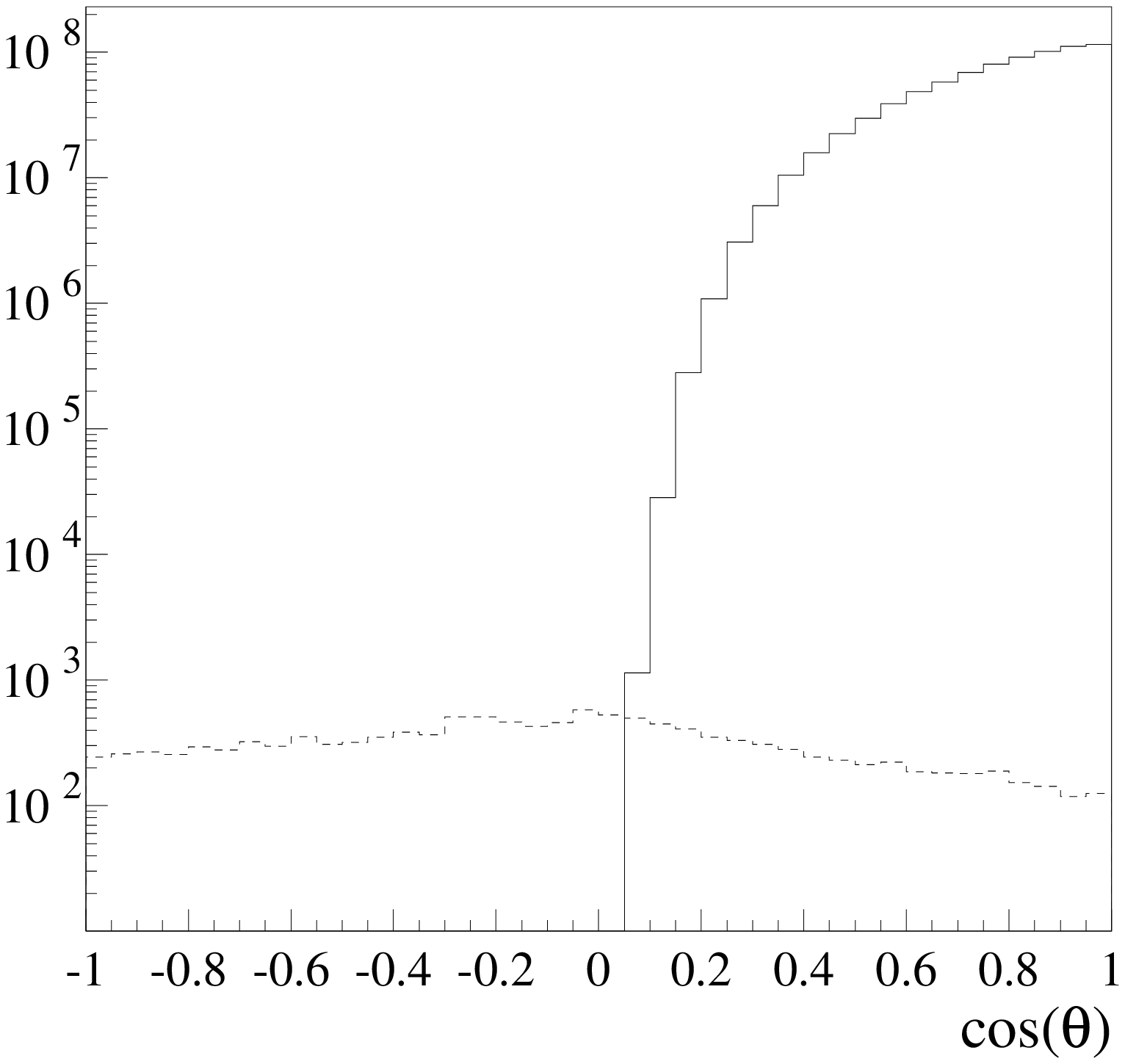,height=6.5cm}}
  \mbox{\epsfig{file=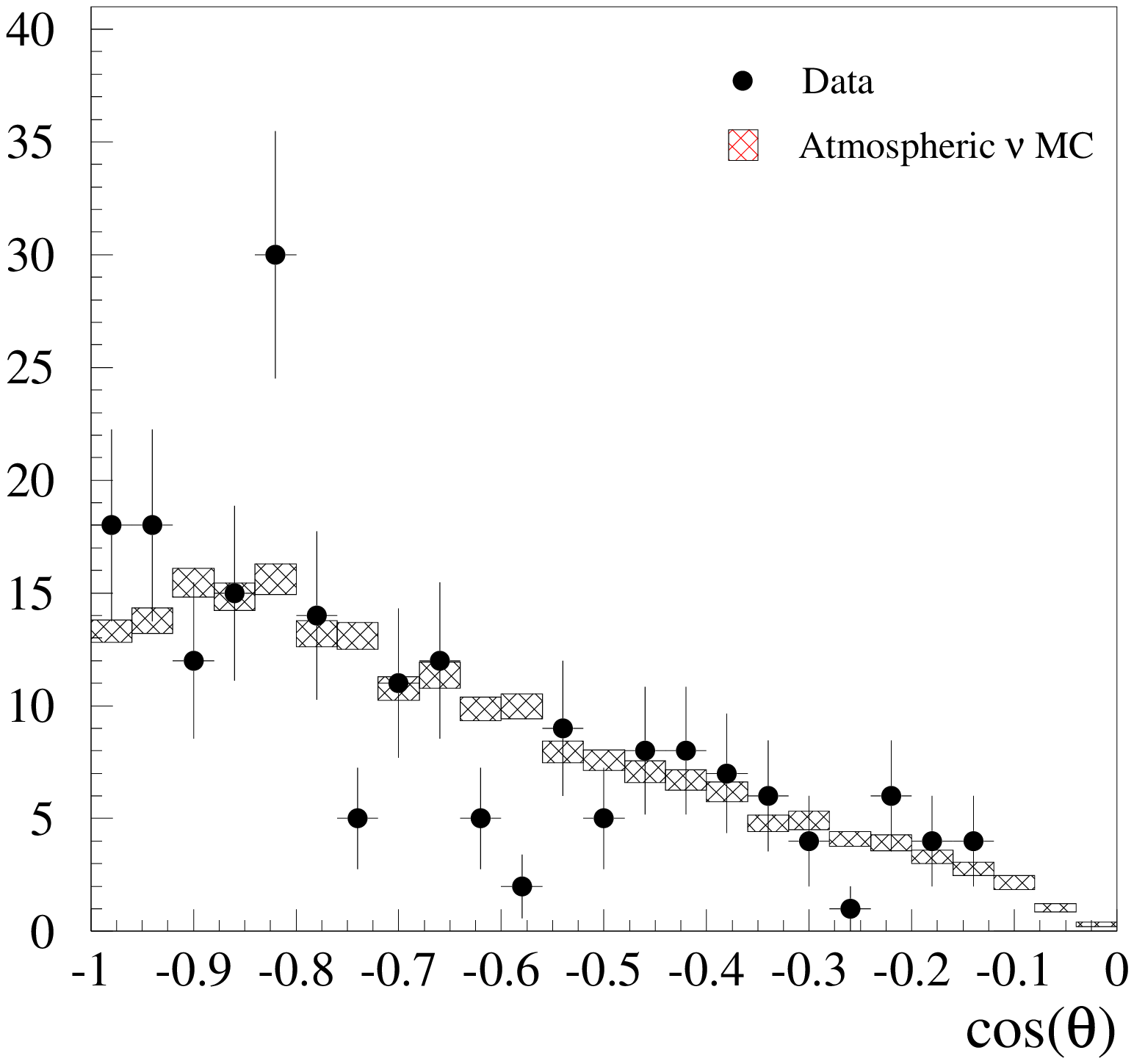,height=6.5cm}}
\caption{Left: The zenith angle distribution of AMANDA triggers.  The solid line
represents triggers from downgoing cosmic ray muons.  The dashed line shows
triggers produced by atmospheric neutrinos.  
Right: The zenith angle distribution of upward reconstructed events. 
The size of the hatched boxes indicates the statistical 
precision of the atmospheric neutrino simulation.  
}
\label{fig:zenith}
\end{figure}

We have searched for atmospheric neutrinos in the 130.1 live days of data 
collected throughout 1997. Calibration and basic detector properties are similar
to the prototype AMANDA-B4 \cite{B4} detector. Figure~\ref{fig:zenith} shows the 
expected trigger level fluxes of downgoing muons ($\sim 6 \cdot 10^6 $ events per day)
and atmospheric neutrinos (a few tens
of events per day). The energy threshold for these atmospheric neutrinos is about  30-50 GeV.

\begin{figure}[htb]
\centering
  \mbox{\epsfig{file=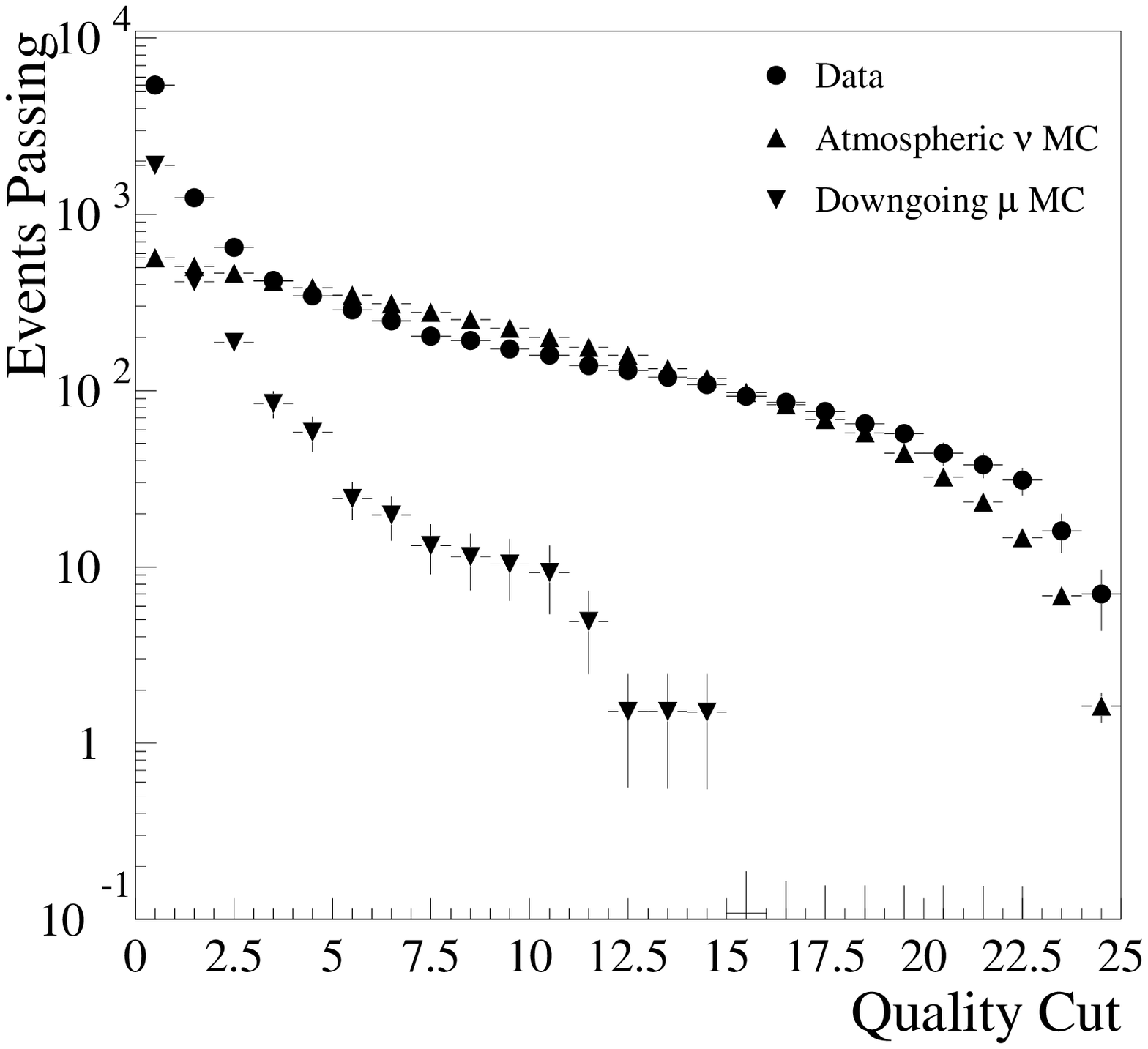,height=6.5cm}}
  \mbox{\epsfig{file=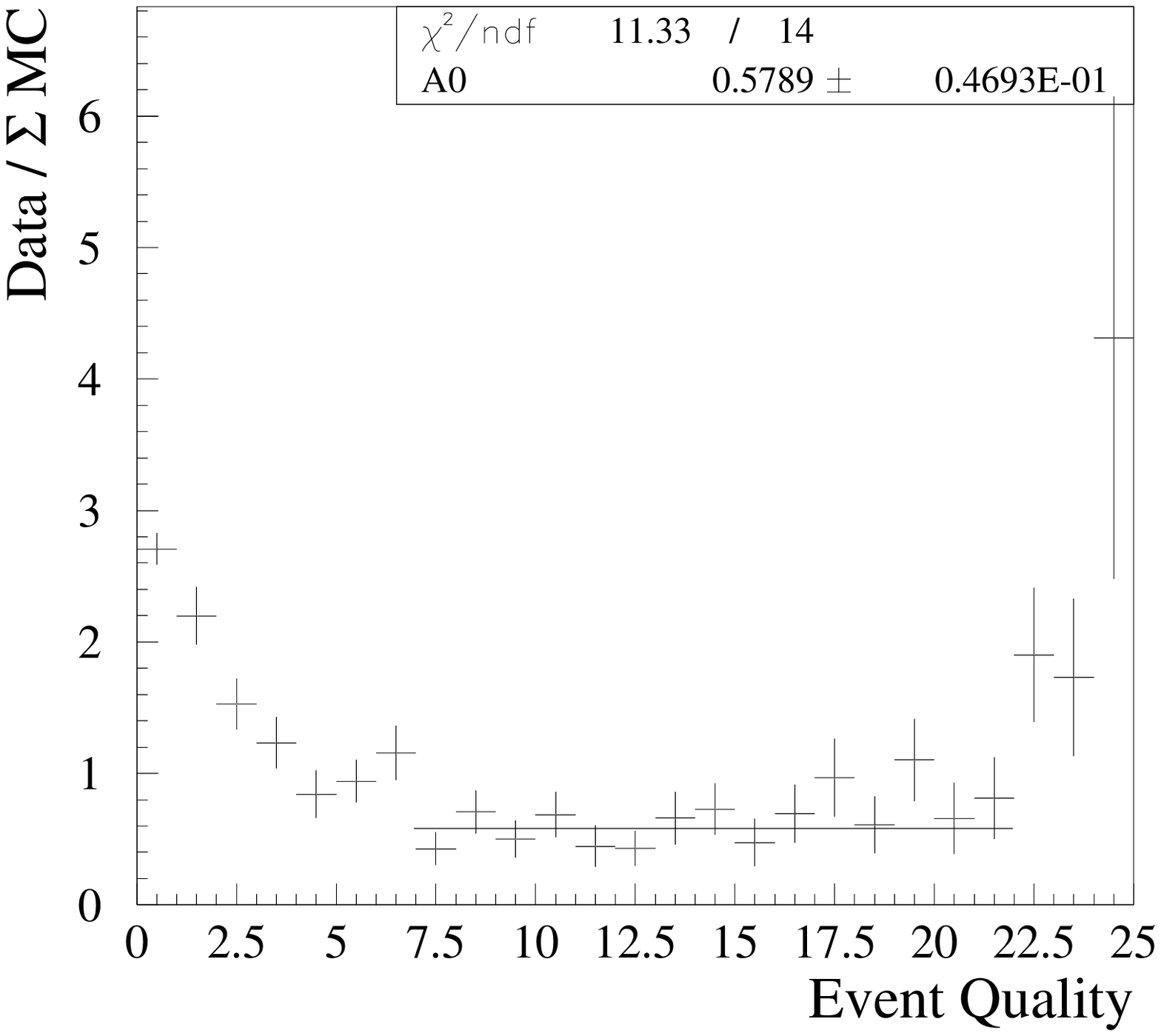,height=6.5cm}}
\caption{Event quality. Left: Passing rates of events above 
a certain quality level is shown for background MC, atmospheric neutrino MC and
experimental data. Right: Differential presentation of the ratio data/MC.}
\label{fig:qual}
\end{figure}

The atmospheric neutrino analysis \cite{nature,icrcatmos} has been performed by two nearly-independent
groups in the collaboration. This has contributed to the introduction of new ideas 
and has provided for some important cross checks of the analysis.
 While we here report
the results of one analysis \cite{ty}, the second \cite{markus,icrcatmos} 
comes to
very similar and statistically consistent conclusions. 
Neutrinos are identified by looking for upward going muons.
We use a maximum likelihood method \cite{wiebusch}, incorporating
a detailed description 
of the scattering and absorption of photons in the ice, 
to reconstruct muon tracks from the measured photon 
arrival times.   
The events in the analysis presented here were  
reconstructed with a Bayesian method \cite{gary}, in which the likelihood function 
is multiplied by a prior probability
function. This follows from  Bayes' theorem, which says that the
probability of an hypothesis given data ($P(H \mid E)$)  is proportional to the product of
the likelihood of the event given the hypothesis ($\mathcal{L}(E \mid H)$), and  the 
prior probability of the hypothesis $P(H)$
\begin{equation}
\label{bayestheorem}
    P(H\mid E) \propto {\mathcal{L}(E \mid H) P(H)}
\end{equation}
We know a priori that downgoing muon hypotheses have a zenith dependence, and
are overall more likely than upgoing hypotheses by about  5 orders of magnitude. We
therefore search for the hypothesis $H$ that maximises
 the joint probability distribution ${\mathcal{L}(E \mid H) P(H)}$.
This greatly reduces   
the number of downgoing muons that
are misreconstructed as upgoing. A small 
fraction of the downgoing muons ($5\cdot10^{-6}$)
are reconstructed as upward and form a background
to the neutrino-induced events.
This background is removed by applying quality criteria 
to the time profiles of the observed photons
as well as to their spatial distribution in the array. 
A measure of the event quality
has been defined by combining six quality variables into a 
single parameter. 
A high event quality is reached when the values 
of all six parameters agree with the 
characteristics of a correctly reconstructed muon track. 
The reconstruction and increasingly stringent cuts
on the event quality reduces the background of a total of  $1.2\cdot 10^9$ events
by a factor of
approximately $10^{8}$, while retaining
about 5\% of the neutrino signal (\ref{table1}.) 
The distribution of the single quality parameter for experimental data and for 
a Monte-Carlo simulation of atmospheric neutrinos 
is shown in Figure~\ref{fig:qual}.
It compares the
number of events passing various levels of cuts; i.e., the integral number
of events above a given quality.  At low qualities, the data set is dominated
by misreconstructed downgoing muons, most of which are reproduced in the
Monte Carlo.  At higher cut levels, the passing rates of data closely track
the simulated neutrino events, and the predicted background contamination
is very low.


We can investigate the agreement between data and Monte Carlo more 
systematically by comparing the differential number of events,
rather than the total number of events passing various
levels of cuts.  This is done in figure~\ref{fig:qual} (right), where the ratios
of the number of events observed to those predicted from the combined
signal and background simulations are shown.  One can see that at low quality
levels there is an excess in the number of misreconstructed events observed.
This is mainly due to instrumental effects such as cross talk which
are not well described in the detector Monte Carlo.  There is also an excess,
though statistically less significant, at very high quality levels, which is
caused by slight inaccuracies in the description of the optical parameters
of the ice.  Nevertheless, over the bulk of the range there is close
agreement between the data and the simulations, apart from an overall
normalization factor. 
In the range where the line is shown the ratio of Data/MC is 
about 0.6. Counting all events 
above the quality cut (7.0) this 
ratio is 0.70. 
It should be emphasized that the 
quality parameter is a combination of all six quality parameters, and so the
flat line in figure~\ref{fig:qual} demonstrates agreement not only in
individual cut parameters but also quantitative agreement in the 
correlations between cut parameters.

\begin{table}
\begin{center}
\begin{tabular}{|l|c|c|c|c|}
\hline
             &  Experimental Data   &  MC: Atmospheric Neutrinos  \\ 
\hline
Triggered    &  $1.2\cdot 10^9$  &    4600       \\ 
\hline
Reconstructed upward    &  $5 \cdot 10^3$  &  571  \\ 
\hline
Upward going  &  $ 204 $       &   279    \\ 
\hline
\end{tabular}
\caption{Event numbers are given at various cutlevels: 
Experimental data and atmnospheric neutrino Monte-Carlo.
}
\label{table1}
\end{center}
\end{table}


The zenith angle distribution for the 204 events is shown 
in Figure \ref{fig:zenith}, and compared to that for the signal 
simulation. In the figure the Monte Carlo events were 
normalized to the observed events. 
The achieved agreement in the 
absolute flux of atmospheric neutrinos is consistent with 
the systematic uncertainties of the absolute sensitivity 
and the flux of high energy atmospheric neutrinos. 
The shape of the zenith distribution of data is statistically consistent with the 
prediction from atmospheric neutrinos.
The zenith distribution 
reflects the angular acceptance of the narrow but tall detector. 
Two hundred and twenty three events were found in an independent analysis \cite{markus}. 
The overlap of 
102 events with the sample presented here is within expectations.  
The observation of atmospheric neutrinos at a rate consistent 
with Monte-Carlo prediction establishes AMANDA-B10 as 
a neutrino telescope.

\section{Neutrino science with AMANDA}
The AMANDA-B10 data has been searched for evidence of several classes of
neutrinos, and for magnetic monopoles. The following sections briefly
describe the status of these searches.

\subsection{Search for a diffuse high energy neutrino flux}

Following on from the observation of atmospheric neutrinos, the first
search we report is that for a diffuse flux of extra-terrestrial neutrinos.

\begin{figure}[htb]
\centering
\raisebox{2 cm}
{
  \mbox{
\epsfig{file=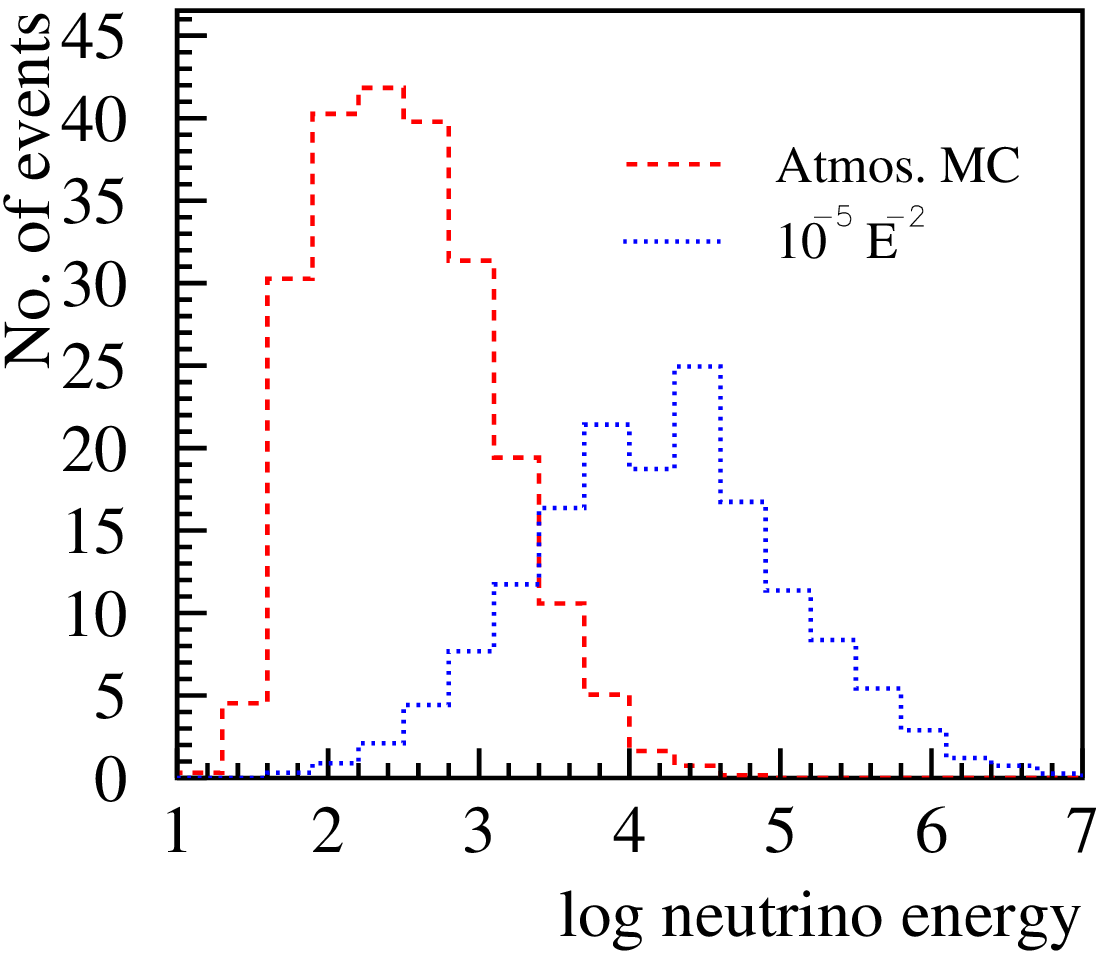,width=6.6cm}}
}
  \mbox{
  \epsfig{file=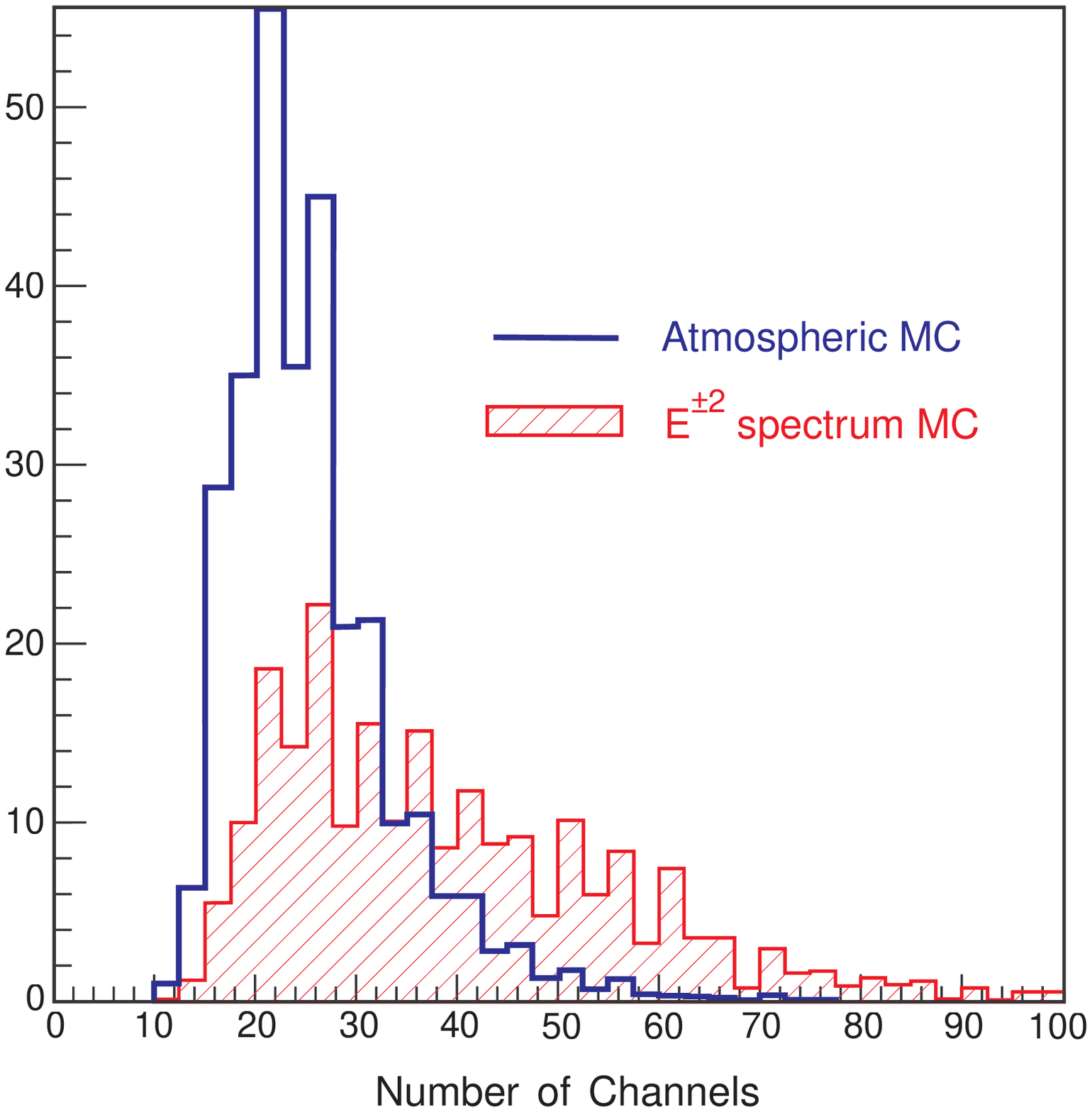,width=7.7cm}}
\vspace{-1.5 cm}
\caption{Left: Monte-Carlo simulation of the energy spectrum of 
atmospheric neutrinos shown in the skyplot in figure ref{fig:sky}.
Also shown is the energy spectrum of neutrinos generated by 
a neutrino flux of a $E_{\nu}^{-2}$-type energy spectrum (see text). 
}
\label{fig:diffuse}
\end{figure}

These neutrinos are expected to have a harder
energy spectra ($\sim E^{-2}$) than that of the atmospheric
neutrino background ($\sim E^{-3.7}$). Some method of energy determination
must therefore
be used. A simple measure of the energy of the muon, and thus indirectly
the energy of the original neutrino, is the number of optical modules that fire
during the passage of the muon. Higher multiplicities correspond to 
higher muon and neutrino energies. 
Figure \ref{fig:diffuse} shows the energy distribution
of events that pass the neutrino filter as predicted for 
a) atmospheric neutrinos and b) an assumed energy spectrum for 
astrophysical neutrinos following a power law of 
$ dN/dE_{\nu} = 10^{-5} E_{\nu}^{-2} \rm\, cm^{-2}\, s^{-1}\, 
sr^{-1}\, GeV^{-1} $. 
When using the number of fired optical modules as a measure of energy we obtain
the distributions given in figure \ref{fig:diffuse}.
The assumed astronomical neutrino flux would 
generate a significant excess at high numbers
of fired optical modules. 
A preliminary analysis does not show such an excess. 
This  leads to a preliminary upper limit \cite{nu2000} ($90\%$C.L.) of 
 $ dN/dE_{\nu}  \approx 10^{-6} E_{\nu}^{-2} \rm\, cm^{-2}\, s^{-1}\, 
sr^{-1}\, GeV^{-1} $.
However, the systematics of this
analysis with respect to the high energy sensitivity 
are still subject to further investigation.
A re-analysis with an updated version of the Monte-Carlo simulation is 
underway.
\vspace{-2cm}
This sensitivity on the diffuse neutrino flux 
is below previously stated upper limits by experiments such as 
Baikal \cite{baikal}, SPS-DUMAND \cite{dumand}, 
AMANDA-A \cite{amanda-a}, and FREJUS \cite{frejus}, 
and comparable to a recently presented  Baikal \cite{baikal01} limit.
It is comparable to the AGN prediction 
by Salamon and Stecker \cite{stecker}
and approaches the prediction of Protheroe \cite{protheroe}. 

\vspace{0.7cm}

\subsection{Point sources}

\begin{figure}[htb]
\centering
\mbox{
     \epsfig{file=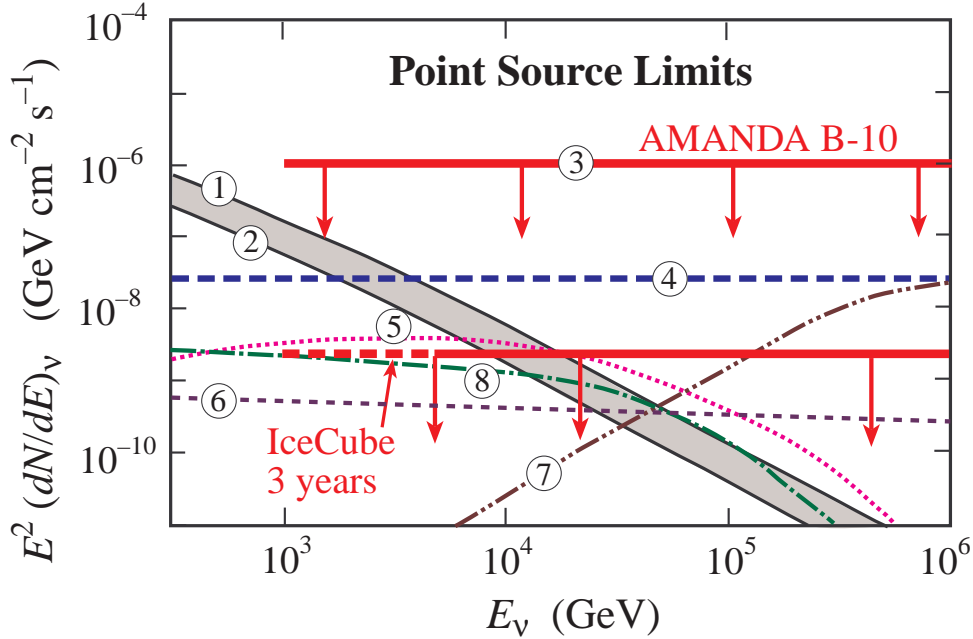,width=13cm}
}
\caption{Limits on high energy neutrinos from point sources.
The preliminary point source limit obtained with AMANDA-B10 (3), as well as the 
sensitivity of the proposed Icecube array, are shown. Curves 1 \& 2 - horizontal
and vertical atmospheric
neutrino background in a $2^{\circ}\times 2^{\circ}$ search bin.  See text for description of models 4--8.
}
\label{fig:point-flux}
\end{figure}

A search for a point source of neutrinos is, in some ways, easier than the diffuse
neutrino flux search. Not only can we dramatically reduce the background by searching
in a small angular bin about the postulated source direction,
 we can directly measure this background
to the search by counting the events in the off-source bins at the same
declination. This is in  contrast to the diffuse search where the background has to
be estimated from Monte Carlo. Of course, determining the sensitivity of the 
detector to calculate an expected signal (for determining a flux limit) must
still rely on Monte-Carlo.  
We have searched for both specific point sources, and also for ``hidden'' point sources,
the latter through an all-sky search. 
The median angular resolution of the AMANDA-B10 array is $3^{\circ}$, giving 319
search bins in an all-sky search. 
In absence of a signal we calculate upper limits 
to a neutrino flux from point sources. 
The preliminary average neutrino flux limits are 
at a level of 
$ dN/dE_{\nu}  \sim  10^{-6} E_{\nu}^{-2} \rm\, cm^{-2}\, s^{-1}\, 
GeV^{-1} $.

One point source search of particular interest is that for Markarian 501. The
AMANDA neutrino flux limit for this source is only about a factor of ten 
above the level of the gamma-ray emission during Markarian 501's
 high gamma-ray emission state in 1997. This means that the AMANDA detector
has reached the level of sensitivity where models that predict neutrino
emissions of the order of the gamma emissions can be tested.

Figure \ref{fig:point-flux} shows the expected neutrino fluxes
from various sources, together with the current preliminary AMANDA 
upper limit ($90\%$ C.L.). The atmospheric neutrino background is 
given for a 
2 by 2 degree bin.
Also indicated in the figure is the expected limit that might
be achieved by the proposed Icecube detector, a kilometre
scale array planned for construction at the south pole. Several models of point
source neutrino production are shown: curve 4 -- 3C273 pp neutrinos \cite{nel}, 5 -- Crab Nebula \cite{crab}, 
6 -- Coma Cluster \cite{coma},
7 -- 3C273 $p\gamma$ neutrinos \cite{stecker}, 8 -- Supernova IC443 \cite{gai}. 

\subsection{Gamma-Ray Bursts}
A search for neutrinos in coincidence with gamma-ray bursts has been conducted.
According to the relativistic fireball model, gamma-ray bursts (GRBs) are
expected to be astrophysical sources of high energy neutrinos \cite{wax,wax2,wax-bah}.
The GRB direction and burst time are known from satellite observations of these
objects. In some models, neutrinos are expected within a short time ($\sim 10$ seconds) 
of the gamma-rays. This allows us to make nearly background free observations, by 
restricting the search bin to a specified direction and time.
With $\sim 1/3$ sky coverage, the BATSE satellite instrument 
detected 304 gamma-ray bursts in 1997. 
AMANDA data for 78 gamma-ray northern hemisphere bursts detected 
on-board the BATSE satellite have been examined for coincident neutrino 
emission. 
No excess of neutrinos has been found above a background of 
17.2 events for all 78 bursts. Due to the low background, the detection
of even a few events from a single burst would be a significant observation.

\subsection{WIMPs}

AMANDA can be used to search for non-baryonic dark matter in the 
form of weakly interacting 
massive particles (WIMPs). 
A promising WIMP candidate, the neutralino, is 
provided by the Minimal Supersymmetric 
extension to the Standard Model of particle physics (MSSM).
Assuming that the dark matter in the Galactic halo is (at least 
partially) composed of relic neutralinos, which were 
formed in the early universe, 
these massive particles do have a probability to get gravitationally trapped in 
the Earth and other massive objects in the Galaxy (sun, galactic center). 
In this theory, the WIMPs lose energy by elastic scattering on nuclei and 
concentrate close to the core of the Earth. There they can annihilate and 
neutrinos can be produced in the decay of the created particles.
Thus, the search for nearly vertical up-going neutrinos 
can be used to constrain the parameter space of supersymmetry. 
No excess of vertical up-going neutrinos has been found. 

The non observation of an excess of vertically up-going 
muons has been used to set a limit on the 
    flux of neutrinos from WIMP annihilations in the center of the Earth \cite{wimp}.
    With only 130 days of exposure in 1997, AMANDA has reached 
    a sensitivity in the region of high WIMP masses ($\geq$ 500\,GeV)  
that begins to constrain the theoretically allowed parameter space. It is 
comparable in sensitivity to other detectors with much longer live-times.

\subsection{Supernova}

By monitoring bursts of low energy neutrinos AMANDA can be used
to detect the gravitational collapse of supernovae in the galaxy.
This method takes advantage of the low noise characteristics (500 - 1500 Hz/PMT) 
of the optical sensors in the deep ice. 
A sensitivity for about $70\%$ of the galaxy is reached at a $90\%$ 
detection efficiency \cite{sn}.

\subsection{Magnetic Monopoles}

A magnetic monopole with unit magnetic Dirac charge and a velocity
of $\beta$ close to 1 would emit Cherenkov light along its path,
exceeding that of of a bare relativistic muon by a factor of 8300.
From the non-observation of events with this clear signature,
a limit of $0.62 \cdot 10^{-16}$ cm$^{-2}$\,s$^{-1}$\,sr$^{-1}$
for highly relativistic monopoles has been derived -- a factor
of 20 below the Parker bound and a factor of four below best
other limits.

\section{Conclusions and Outlook}
 
The detection of atmospheric neutrinos in agreement with expectation 
 establishes AMANDA as a neutrino telescope. 
Since February 2000, the significantly larger and improved 
AMANDA-II array has been collecting data.
Its effective area
for high energy neutrinos is about three times that of
 the B10 array. At the same time improved angular resolution 
and background rejection potential are available.
The analysis of these data is under way and will improve the given results 
significantly. 
A proposal exists to construct the Icecube detector which would consist
of 4800 photomultipliers to be deployed on 80 strings. 
It will allow us to reach $\sim 1\,{\mathrm km^2}$ effective telescope
area, above an energy of 1\,TeV with an angular resolution of
well below 1 degree.

\section*{Acknowledgements}

  This research was supported by the U.S. NSF office of Polar Programs
and Physics Division, the U. of Wisconsin Alumni Research Foundation,
the U.S. DoE, the Swedish Natural Science Research Council, the
Swedish Polar Research Secretariat, the Knut and Alice Wallenberg
Foundation, Sweden, the German Ministry for Education and Research,
the US National Energy Research Scientific Computing Center (supported
by the U.S. DoE), U.C.-Irvine AENEAS Supercomputer Facility, and
Deutsche  For-schungsgemeinschaft (DFG).  D.F.C. acknowledges the
support of the NSF CAREER program.  P. Desiati was supported by the
Koerber Foundation (Germany).  C.P.H. received support from the EU 4th
framework of Training and Mobility of Researchers.  St. H. is
supported by the DFG (Germany).  P. Loaiza was supported by a grant
from the Swedish STINT program.

\end{document}